\documentclass[aps,prb,amsmath,twocolumn,amssymb,titlepage,superscriptaddress,showpacs]{revtex4-1}

\usepackage{soul}
\usepackage{epsfig}% Include figure files
\usepackage{graphicx}% Include figure files
\usepackage{lipsum}% http://ctan.org/pkg/lipsum
\usepackage{dcolumn}% Align table columns on decimal point
\usepackage{bm}% bold math
\usepackage[utf8]{inputenc}
\usepackage{graphicx} 
\usepackage{xcolor}
\usepackage{float}

\newcommand{\astcycl}{\mathrlap{\kern0.085em{\circlearrowright}}\ast}
\newcommand{\taucycl}{\mathrlap{\kern0.42em{\bullet}}\circlearrowright}

\setulcolor{red}
\setul{0.3ex}{0.4ex}
\begin{document}

\title{Uncovering the nature of transient and metastable non-equilibrium phases in 1$T$-TaS$_2$}

%Identifying the transient and metastable non-equilibrium phases in the Peierls-Mott insulator 1$T$-TaS$_2$  }
% Uncovering the nature of transient and metastable nonequilibrium phases in 1$T$-TaS$_2$ 
%\title{Evolution of the electronic band structure with Sulfur-doping in Ta$_2$NiSe$_5$  and its three-dimensionality revealed by angle-resolved photoemission}

\author{Tanusree Saha }
\altaffiliation{Corresponding author: tanusree.saha@student.ung.si}
\affiliation{Laboratory of Quantum Optics, University of Nova Gorica, 5270 Ajdov\v{s}\v{c}ina, Slovenia}

\author{Arindam Pramanik }
%\altaffiliation{Corresponding author: tanusree.saha@student.ung.si}
\affiliation{Department of Theoretical Physics, Tata Institute of Fundamental Research, Mumbai 400005, India}

\author{Barbara Ressel}
\affiliation{Laboratory of Quantum Optics, University of Nova Gorica, 5270 Ajdov\v{s}\v{c}ina, Slovenia}

\author{Alessandra Ciavardini}
\affiliation{Laboratory of Quantum Optics, University of Nova Gorica, 5270 Ajdov\v{s}\v{c}ina, Slovenia}
%\affiliation{Elettra Sincrotrone Trieste, Strada Statale 14 km 163.5, 34149 Trieste, Italy}

\author{Fabio Frassetto}
\affiliation{CNR-Institute of Photonics and Nanotechnologies (CNR-IFN), 35131 Padova, Italy}

\author{Federico Galdenzi}
\affiliation{Laboratory of Quantum Optics, University of Nova Gorica, 5270 Ajdov\v{s}\v{c}ina, Slovenia}
%\affiliation{Elettra Sincrotrone Trieste, Strada Statale 14 km 163.5, 34149 Trieste, Italy}

\author{Luca Poletto}
\affiliation{CNR-Institute of Photonics and Nanotechnologies (CNR-IFN), 35131 Padova, Italy}

\author{Arun Ravindran}
\affiliation{Laboratory of Quantum Optics, University of Nova Gorica, 5270 Ajdov\v{s}\v{c}ina, Slovenia}
%\affiliation{Elettra Sincrotrone Trieste, Strada Statale 14 km 163.5, 34149 Trieste, Italy}

\author{Primo\v{z} Rebernik Ribi\v{c}}
%\affiliation{Laboratory of Quantum Optics, University of Nova Gorica, 5001 Nova Gorica, Slovenia.}
\affiliation{Elettra Sincrotrone Trieste, Strada Statale 14 km 163.5, 34149 Trieste, Italy}

\author{Giovanni De Ninno}
\affiliation{Laboratory of Quantum Optics, University of Nova Gorica, 5270 Ajdov\v{s}\v{c}ina, Slovenia}
\affiliation{Elettra Sincrotrone Trieste, Strada Statale 14 km 163.5, 34149 Trieste, Italy}

\begin{abstract}

Complex systems are characterized by strong coupling between different microscopic degrees of freedom. Photoexcitation of such materials can drive them into new transient and long-lived hidden phases that may not have any counterparts in equilibrium.  By exploiting femtosecond time- and
angle-resolved photoemission spectroscopy, we probe the photoinduced transient phase and the recovery dynamics of the ground state in a complex material: the charge density wave (CDW)-Mott insulator 1$T$-TaS$_2$. We reveal striking similarities between the band structures of the transient phase and the (equilibrium) structurally undistorted metallic phase, with evidence for the coexistence of the low-temperature Mott insulating phase and high-temperature metallic phase. Following the transient phase, we find that the restoration of the Mott and CDW order begins around the same time. This highlights that the Mott transition is tied to the CDW structural distortion, although earlier studies have shown that the collapse of Mott and CDW phases are decoupled from each other. Interestingly, as the suppressed order starts to recover, a long-lived metastable phase emerges before the material recovers to the ground state. Our results demonstrate that it is the CDW lattice order that drives the material into this metastable phase, which is indeed a commensurate CDW-Mott insulating phase but with a smaller CDW amplitude. Moreover, we find that the long-lived state emerges only under strong photoexcitation and has no evidence when the photoexcitation strength is weak. 

%where the Mott electronic order and CDW lattice order are suppressed,
\end{abstract}
\date{\today}

\maketitle

\section{INTRODUCTION}

Materials dominated by strong electron-electron and electron-lattice interactions can undergo phase transitions to insulating ground states, exhibiting charge and lattice order\cite{ch6_mott1961,ch6_mott1968,ch6_gruner1994,ch6_imada1998,ch6_knox1964,ch6_zandt1974,ch6_gebhard2000}. Under non-equilibrium conditions, such systems display a collapse of charge and lattice order of the ground state, as well as the occurrence of novel or hidden phases which are thermally inaccessible under equilibrium\cite{ch6_ichikawa2011,ch6_stojchevska2014}. Ultrafast pump-probe techniques have paved the way to delve into the non-equilibrium regime of matter\cite{ch6_hilton2006,ch6_bovensiepen2012,ch6_miller2014}. Solid-state systems exhibiting some intriguing phases, such as Mott\cite{ch6_V2O3_1983,ch6_boer1937,ch6_limelette2003,ch6_janod2015,ch6_katsufuji1994,ch6_ma1993}, charge density wave (CDW)\cite{ch6_wang2003,ch6_moncton1975,ch6_boubeche2021,ch6_blackburn2013,ch6_hashimoto2014} and excitonic\cite{ch6_cercellier2007,ch6_bucher1991,ch6_du2017,ch6_lu2017},
are being extensively studied using ultrafast spectroscopic and diffraction methods in the femtosecond time domain. The relevant timescales of quenching dynamics, photoinduced phase transitions\cite{ch6_wall2011,ch6_rohwer2011,ch6_smallwood2011,ch6_hellmann2012,ch6_peterson2011,
ch6_tomeljak2009,ch6_schmitt2011,ch6_okomoto2011} and the emergence of metastable phases\cite{ch6_stojchevska2014,ch6_vaskivskyi2015,ch6_sun2018,ch6_shi2019} are the topics of great interest. While the quenching occurs instantaneously in Mott insulators and the timescale is set by the electronic hopping time given by the bandwidth\cite{ch6_wall2011,ch6_moritz2010}, the Peierls-CDW materials exhibit quenching times that are comparable to the timescales of the, slower, lattice-driven processes\cite{ch6_tomeljak2009,ch6_schmitt2008}. For excitonic insulators, carrier screening time, given by the plasma frequency, determines the characteristic timescale\cite{ch6_rohwer2011}. %Majority of the studies have focussed on the early stages of the dynamics, i.e., on the collapse rather than the recovery of the ground state. In this Letter, we address the above scenario by tracking the recovery dynamics of electronic and lattice order, alongside  in a complex system, 1$T$-TaS$_2$, employing angle-resolved photoemission spectroscopy \cite{damascelli2003} (ARPES) in the ultrafast time domain. 

\begin{figure}[t]
\centering
%\vspace{-4ex}
\includegraphics[width = 1\linewidth]{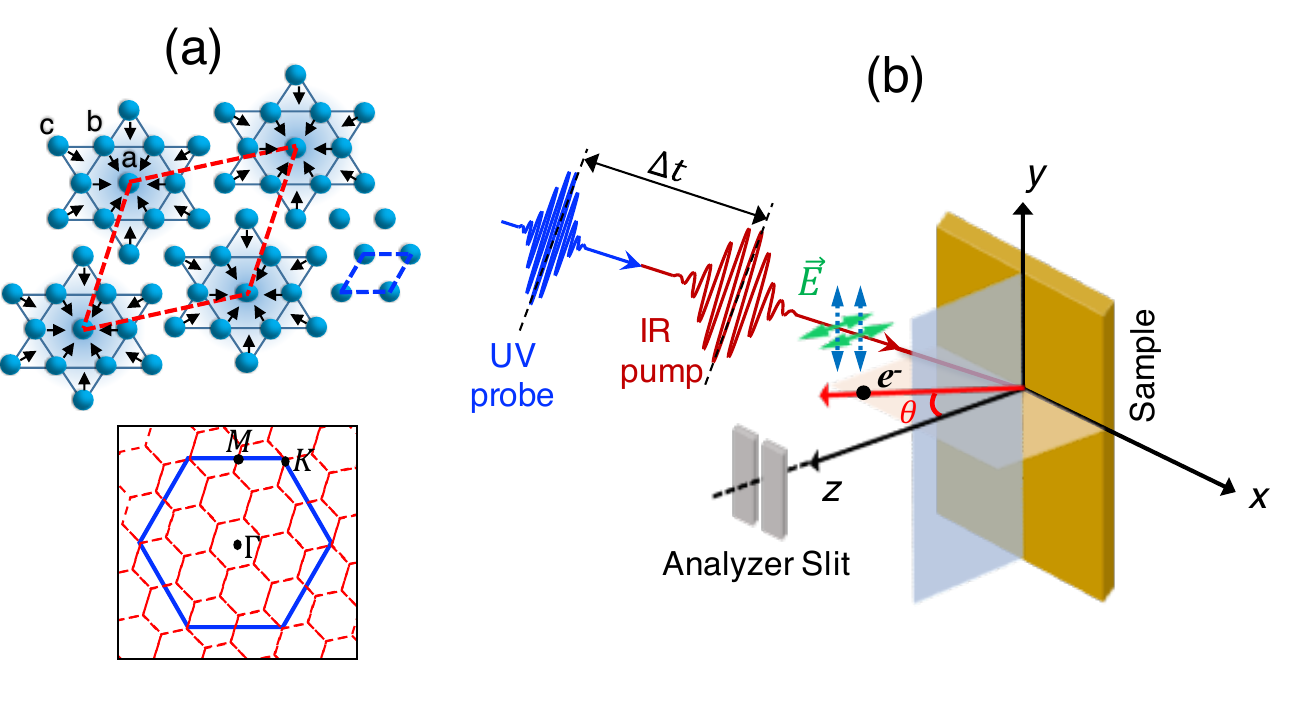}
\caption{(a) (Top) In-plane structural distortion in the CCDW phase of 1$T$-TaS$_2$ produces ``Star-of-David" clusters having inequivalent ``a", ``b", and ``c" atoms. Red and blue dashed lines indicate the unit cells in the CCDW and unreconstructed phases, respectively. The arrows indicate the displacement of the Ta atoms from their initial positions. (Bottom) Brillouin zone in the unreconstructed (blue) and distorted phases (red) with the high-symmetry points $\Gamma, M, K$. (b) A schematic of the pump-probe experimental geometry where the electric field $\vec{E}$ of s- and p-polarized photons are indicated by blue (along \textit{y}-axis) and green (in the \textit{xz}-plane) double-headed arrows, respectively. }
\label{Figure1ab}
%\vspace{-2ex}
\end{figure}

The layered CDW-Mott insulator 1$T$-TaS$_2$ is a prominent example of complex system since both electron-electron and electron-lattice interactions are simultaneously strong. It exhibits a manifold of electronic and structurally ordered phases\cite{ch6_hellmann2012,ch6_smith2006,ch6_perfetti2006,ch6_perfetti2008,ch6_sohrt2014,ch6_wang2020}: at high temperatures ($T>$ 550 K), the system has an undistorted hexagonal structure and is metallic while cooling results in the formation of various CDW phases - incommensurate $\rightarrow$ nearly-commensurate $\rightarrow$ commensurate. Below the critical temperature for the commensurate CDW (CCDW) phase, $T_C=180$ K, a periodic lattice distortion (PLD) gives rise to the formation of ``Star-of-David (SD)'' - shaped clusters consisting of thirteen Ta atoms. Fig.~\ref{Figure1ab}(a) shows a schematic of the lattice reconstruction in the plane of Ta atoms and its Brillouin zone in the metallic and CCDW phases of 1$T$-TaS$_2$. The $\sqrt{13}\times\sqrt{13}$ superlattice splits the Ta $5d$ valence band into three subband manifolds, such that the narrow half-filled band at the Fermi level $E_F$  becomes favorable for a Mott-Hubbard transition\cite{ch6_smith2006,ch6_fazekas1979}. Previous trARPES studies have shown an instantaneous collapse of the Mott gap at $E_F$ on timescales $<50$ fs after photoexcitation\cite{ch6_hellmann2012,ch6_peterson2011,ch6_sohrt2014,ch6_ligges2018}. In addition, the CDW gap between the Ta $5d$ subbands was found to melt faster than the lattice vibrational timescale, suggesting that electron correlations might play a vital role in the CDW ordering\cite{ch6_peterson2011,ch6_sohrt2014}. A prompt collapse of charge ordering was also shown using ultrafast core-level PES\cite{ch6_hellmann2010}. Ultrafast electron diffraction studies have identified a suppression of the PLD in the nearly-CCDW phase from the optically induced change in the spatial distribution of the electron density\cite{ch6_eichberger2010}. Lately, single-shot time-resolved techniques were able to capture the emergence of a persistent ``hidden'' phase in 1$T$-TaS$_2$\cite{ch6_stojchevska2014,ch6_vaskivskyi2015,ch6_sun2018,ch6_vaskivskyi2016,ch6_gerasimenko2019,ch6_stahl2020,ch6_ravnik2021,
ch6_gao2022}. However, different characteristics of such a state can be manifested by tuning the experimental conditions\cite{ch6_avigo2016}.

%\vspace{-8ex}

Even though this material has been extensively studied, there has been minimal emphasis on the state of charge and lattice ordering in the non-equilibrium transient phase. Moreover, the majority of the studies have focussed on the early stages of the dynamics, i.e., on the collapse rather than the recovery to the ground state. In the present work, we address the above scenario in 1$T$-TaS$_2$ by studying its electronic band structure in the transient phase, as well as the recovery dynamics of the electronic and lattice order. We choose band structure as the spectroscopic parameter since its various features such as bandwidth, dispersion of the band and binding energy provide information about the lattice order, which plays a prominent role in the ground state of 1$T$-TaS$_2$. Angle-resolved photoemission spectroscopy\cite{ch6_damascelli2003} (ARPES) in the ultrafast time domain is employed to systematically track the temporal evolution of Ta $5d$ subbands in the CCDW-Mott phase. Our time-resolved ARPES (trARPES) study demonstrates that, after optical excitation, the material enters a transient phase which bears a striking correspondence with the high-temperature unreconstructed phase. Simultaneously, the early dynamics of the photoexcited system also demonstrates the coexistence of Mott-insulating and unreconstructed metallic phases. Interestingly, the recovery of the Mott and CDW dynamics, after traversing the transient phase, is observed to commence around the same time. It is important to note that although the suppression of Mott-CDW electronic order and CDW lattice order is known to occur on two distinct time scales in 1$T$-TaS$_2$\cite{ch6_peterson2011,ch6_sohrt2014}, the presence of a single timescale observed for the order re-establishment emphasizes that Mott physics is indeed coupled to the CCDW ordering in this material. Moving further, we find that the material recovers to a long-lived hidden phase that is primarily governed by the lattice order of the CDW. Moreover, our results predict that the hidden phase is a CCDW-Mott insulating phase but with a reduced CDW amplitude. Lastly, we also demonstrate that the emergence of a long-lived metastable state is observed only at high photoexcitation strengths and has no signatures under weak photoexcitation. 

\begin{figure}
\centering
%\vspace{-4ex}
%\includegraphics[width = 1\linewidth]{Figure1ab.pdf}
\includegraphics[width = 1\linewidth]{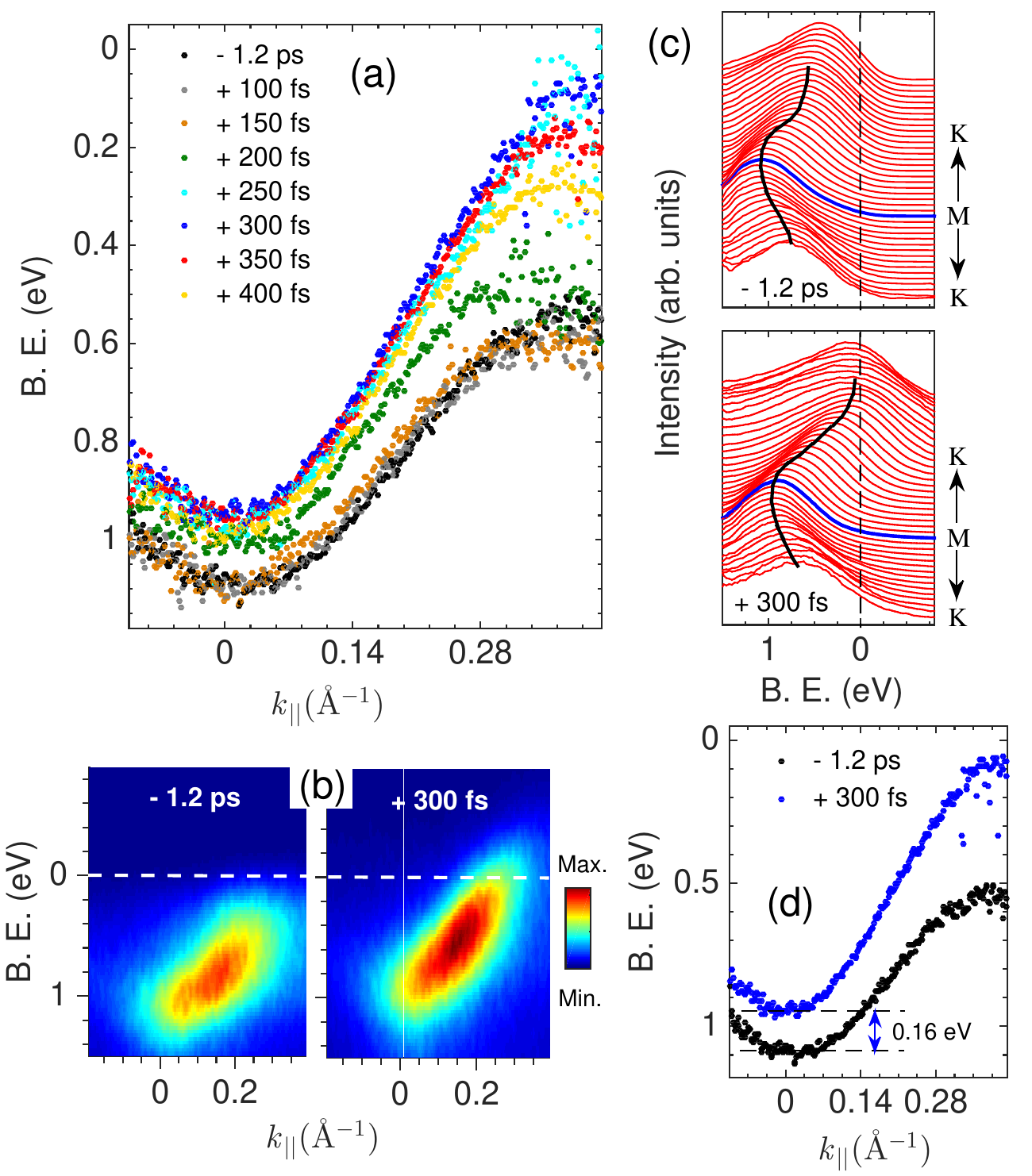}

%\vspace{-2ex}
%\vspace{-2ex}
\caption{(a) Time evolution of the electronic band structure in 1$T$-TaS$_2$ about $M$-point (along $MK$ direction). The peak positions of the energy distribution curves (EDCs) have been plotted as a function of $k_{||}$ at each pump-probe delay $\Delta t$. (b) ARPES snapshots acquired before and after ($\Delta t$ = +300 fs) photoexcitation. (c) Corresponding EDC stacking where the blue curve represents the EDC at $M$. The black curves are guide to the eye for the band dispersion. (d) Comparison of the band dispersion before photoexcitation and in the transient state of the system, where there is an energy shift towards $E_F$ and the band is more dispersive. All the data correspond to a high pump fluence of 3.6 mJ/cm$^2$ and the dashed lines in (b) and (c) indicate $E_F$. Binding energy is abbreviated to B. E.}
\label{Figure1_PP}
%\vspace{-2ex}
\end{figure}
%\vspace{-8ex}

 \begin{figure}[t]
\centering
%\vspace{-4ex}
\includegraphics[width = 1\linewidth]{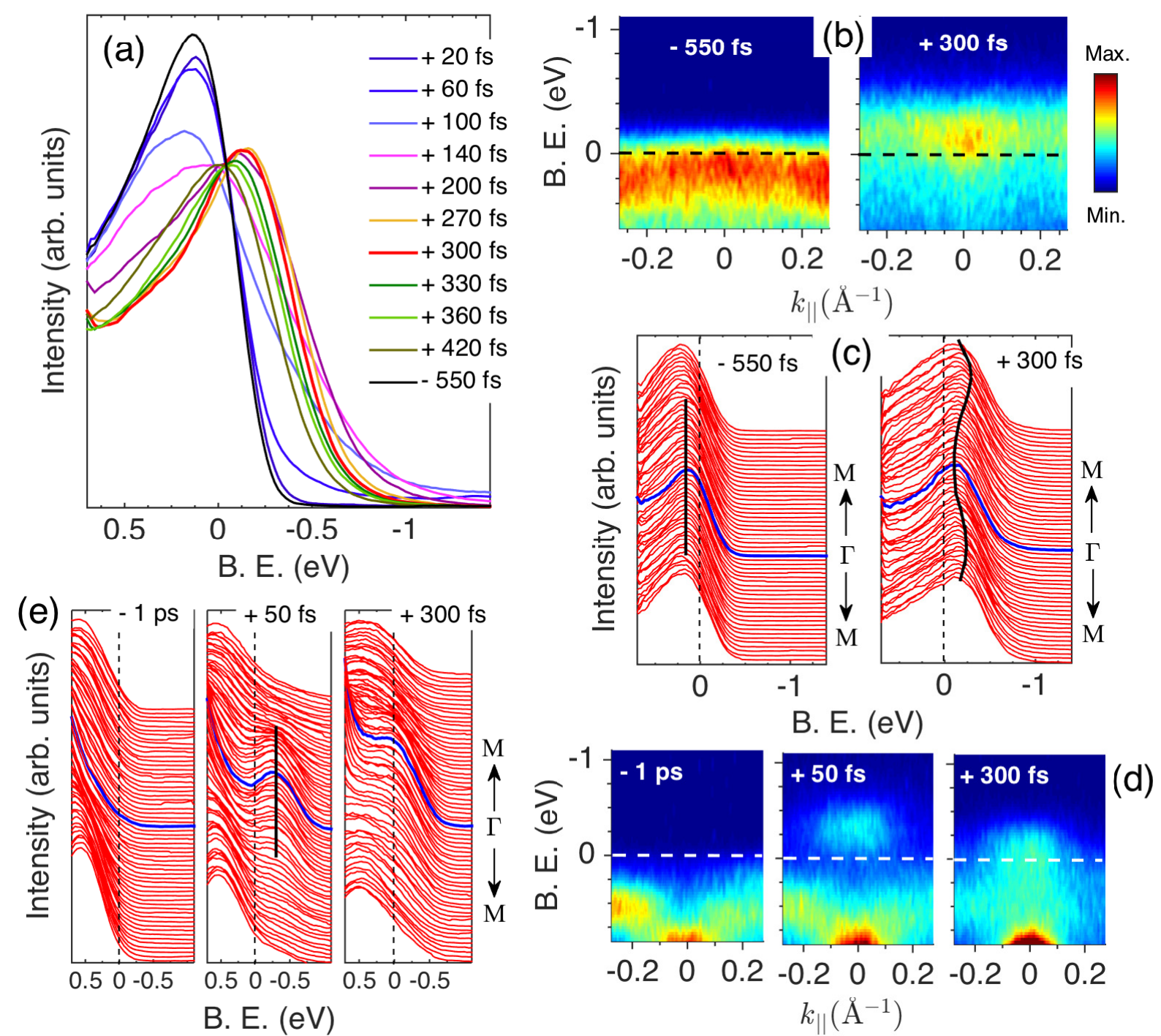}

%\vspace{-2ex}
%\vspace{-2ex}
\caption{(a) Temporal evolution of the EDCs at early pump-probe delays integrated over a $k_{||}$ range of $\pm 0.1$ \AA$^{-1}$ about $\Gamma$-point (along $\Gamma M$ direction). (b) ARPES snapshots acquired before and after ($\Delta t$ = +300 fs) photoexcitation. (c) Corresponding EDC stacking where the blue curve denotes the EDC at $\Gamma$. Smooth curves are guide to the eye to emphasize the change in the band dispersion around $\Gamma$ in the transient phase. (d) ARPES snapshots about $\Gamma$ taken at different delays using s-polarized probe pulses. (e) Corresponding EDC stacking. The smooth black line indicates the flat upper Hubbard band and its dynamics is obtained by changing the probe polarization from horizontal (p-pol) to vertical (s-pol).  The data acquired using p-polarized  and s-polarized probe pulses correspond to pump fluences of 3.2 mJ/cm$^2$ and 3.6 mJ/cm$^2$, respectively, and the dashed lines indicate $E_F$. Binding energy is abbreviated to B. E.}
\label{Figure3_PP_3}
%\vspace{-2ex}
\end{figure}

\section{EXPERIMENTAL DETAILS}
Single crystals of 1$T$-TaS$_2$ were purchased from HQ Graphene\cite{ch6_hqgraphene}. The trARPES experiments were performed at the CITIUS high-harmonic generation (HHG) light source\cite{ch6_citius}. The system is driven by a mode-locked Ti:Sapphire laser delivering 800-nm pulses, with a duration of 40 fs at a repetition rate of 5 kHz. The driving laser was split in two beams: the major part of the intensity was used to generate extreme-ultraviolet (EUV) probe pulses through HHG, with Ar as the generating medium, and the remaining part was used as the pump. The intensity of the pump pulses on the sample was controlled with a variable attenuator - in all experimental plots, the fluence refers to the (incident) peak energy density (in mJ/cm$^2$), determined from the expression $2E_{p}/(\pi w^2)$, where $E_p$ is the  energy per pulse and $w$ is the beam waist at the sample position. A schematic of the experimental geometry showing the polarization of pulses is shown in Fig.~\ref{Figure1ab}(b). The photon energy of the probe was selected by a monochromator grating with off-plane geometry, which preserved the pulse duration\cite{ch6_poletto}. During the experiments, the fundamental frequency of the laser ($h\nu=1.55$ eV) was used for optical excitation (pump pulse). A photon energy $h\nu\sim$ 20 eV (harmonic 13 of the fundamental laser) was selected for the probe pulse due to higher photoionization cross-section of the Ta 5$d$ bands and a high photon flux. To preserve the ultrafast response, the energy resolution of the source was limited to about 150 meV. This allowed us to achieve a temporal resolution of around 50 fs. The ultra-high vaccum setup at CITIUS is equipped with an R3000 hemispherical electron analyser from VG Scienta. A closed-cycle Helium cryostat was used to control the sample temperature and all the measurements were performed at an initial sample temperature $T$ = 100 K. Prior to ARPES measurements, clean sample surfaces were obtained via cleaving in the direction perpendicular to the atomic planes. The samples were cleaved under UHV pressure better than  6x10$^{-9}$ mbar and the measurements were performed at a base pressure $<$ 1x10$^{-10}$ mbar. p-polarized pump and probe pulses [green arrows in Fig.~\ref{Figure1ab}(b)] were used for the obtained data, unless specified.

 %All the measurements were performed at an initial sample temperature $T=100$ K ($<T_C$) and a base pressure $<$ 1x10$^{-10}$ mbar. p-polarized pump and probe pulses [green arrows in Fig.~\ref{Figure1ab}(b)] were used for the obtained data, unless specified.

\section{RESULTS}

We will refer to the (equilibrium) electronic band structure of 1$T$-TaS$_2$ reported in Ref. [44] while presenting the trARPES results on different Ta $5d$ subbands. Firstly, we
will demonstrate the nature of the photoinduced transient phase by characterizing the evolved band structure. For a high photoexcitation strength (3.6 mJ/cm$^2$), the time evolution of the Ta $5d$ subband along high symmetry $MK$ direction (we call it $B_2$ band)\cite{ch6_sohrt2014} is plotted in Fig.~\ref{Figure1_PP}(a). We observe that a shift in binding energy towards $E_F$ and an enhancement of the bandwidth characterize the evolution, which occurs on a 200 fs-timescale. Since the timescale corresponds to half an oscillation cycle of the CDW amplitude mode \cite{ch6_duffey1976,ch6_demsar2002}, the temporal changes indicate the collapse of CDW lattice order after photoexcitation. Subsequent recovery of the suppressed order is observed to occur after 300 fs (red and yellow circles). The characterization of the transient phase at pump-probe delay $\Delta t=+300$ fs is presented in Figs.~\ref{Figure1_PP}(b)-(d). An energy shift of the band minimum by 0.16 eV towards $E_F$, accompanied by a substantial increase of the bandwidth [see Fig.~\ref{Figure1_PP}(d)], are in excellent agreement with the dispersion of the $B_2$ band in the unreconstructed phase\cite{ch6_sohrt2014}. According to theoretical calculations\cite{ch6_smith2006,ch6_sohrt2014}, the dispersion crosses $E_F$ at $k_{||}$ away from $M$, which is however, not evident in our data at 300 fs. This is because $B_2$ might have traversed such a feature within a few 10s of fs before 300 fs and could not be captured due to a large time interval (50 fs) used in the experiments. This particular characteristic of the dispersion is reported in Ref. [31]. Despite the correspondence of the transient band dispersion with that of the (equilibrium) high-temperature phase, the evolved band structure does not reflect phase transitions due to the rise in effective lattice temperature. This is because the observed changes occur much faster than the time scale needed to transfer the energy from the electronic subsystem to the lattice through phonon emission. According to the partial density of states in 1$T$-TaS$_2$ \cite{ch6_smith2006}, photoexcitation involves a redistribution of the conduction electron density within the SD clusters. This results in a radial motion of the Ta atoms towards the outer ring of the SD clusters [``c'' atom in Fig. \ref{Figure1ab}(a)] and hence, a relaxation of the periodic lattice distortion. The electrons can accommodate instantaneously to the atomic positions (Born-Oppenheimer approximation) which is evidenced by the band structures obtained at different time delays in Fig. \ref{Figure1_PP}(a). Hence, the relaxation of the PLD demonstrated in our results is driven by the redistribution of charge density and is not an effect related to the increase in lattice temperature.

\begin{figure}[t]
\centering
%\vspace{-4ex}
\includegraphics[width = 1\linewidth]{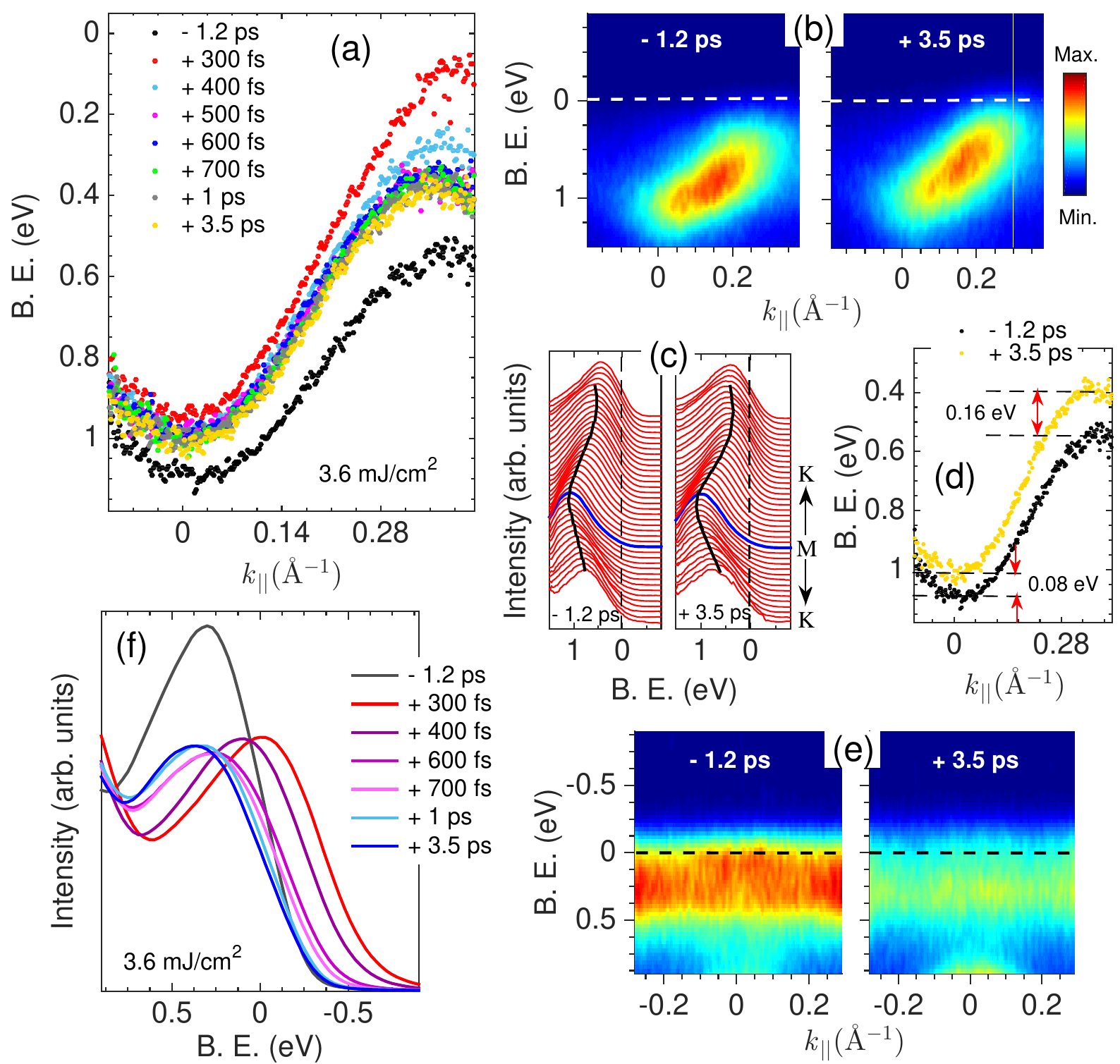}

%\vspace{-2ex}
%\vspace{-2ex}
\caption{(a) Time evolution of the electronic band dispersion about $M$-point (along $MK$ direction). For each pump-probe delay $\Delta t$, the peak position of the EDCs are plotted as a function of $k_{||}$. (b) ARPES snapshots acquired before and after ($\Delta t$ = +3.5 ps) photoexcitation. (c) Corresponding EDC stacking where the blue curve represents the EDC at $M$. The black curves are guide to the eye for the band dispersion. (d) Comparison of the band dispersion before photoexcitation and in the long-lived state of the system. The energy shifts around the band minimum and maximum are indicated by arrows. (e) ARPES snapshots acquired before and after ($\Delta t$ = 3.5 ps) photoexcitation about $\Gamma$-point (along $\Gamma M$ direction). (f) Temporal evolution of the EDCs at longer delays integrated over a $k_{||}$ range of $\pm 0.1$ \AA$^{-1}$ about $\Gamma$. All the data correspond to a high pump fluence of 3.6 mJ/cm$^2$ and the dashed lines in (b), (c), (e) denote $E_F$. Binding energy is abbreviated to B. E.  }
\label{Figure2_PP}
%\vspace{-2ex}
\end{figure}
 
We will now look at the dynamics of the lower Hubbard band (LHB)\cite{ch6_sohrt2014} along high symmetry $\Gamma M$ direction at a similar photoexcitation strength (3.2 mJ/cm$^2$). The EDCs at various time delays extracted from the $k$-integrated trARPES spectrum are shown in Fig.~\ref{Figure3_PP_3}(a). The early dynamics show a collapse of the Mott phase as the spectral weight in LHB is suppressed and transferred to binding energies at and above $E_F$, similar to earlier studies\cite{ch6_perfetti2006,ch6_peterson2011,ch6_hellmann2012}. The recovery of the spectral weight begins after 300 fs; it is to be noted that this is the same time at which the CDW lattice order starts to reform in Fig.~\ref{Figure1_PP}(a). In spite of the established scenario where the suppression of electronic and lattice order occurs on different time scales\cite{ch6_peterson2011}, we find that the re-establishment of Mott electronic order and CDW lattice order begins at the same time. This provides evidence that the CCDW lattice reconstruction is the mechanism behind the Mott transition in this material\cite{ch6_sohrt2014,ch6_smith2006}. Figures~\ref{Figure3_PP_3}(b)-(e) display the characteristics of the band structure in the transient phase at $\Delta t=+300$ fs. We find that the spectral weight from LHB has shifted to an energy band above $E_F$, which is (i) dispersive about $\Gamma$ unlike the flatness of LHB, (ii) the band minima lies at $\approx$ -0.1 eV [see Figs~\ref{Figure3_PP_3}(b), (c)]. [It is to be noted that the dispersive feature beyond $\pm 0.15$\AA$^{-1}$ in Figs. \ref{Figure3_PP_3}(b), (c) (left panels) is a contribution from other Ta $5d$ subbands.] More importantly, the dispersive band at 300 fs does not correspond to the flat upper Hubbard band (UHB). This has been verified from the UHB dynamics that could be tracked at 20 eV probe energy by changing the polarization of the probe pulses [see Fig.~\ref{Figure1ab}(b)] from horizontal (p-pol) to vertical (s-pol). In Figs.~\ref{Figure3_PP_3}(d) and \ref{Figure3_PP_3}(e), the UHB lying at $\approx$ -0.25 eV can be distinctly observed at $\Delta t=+50$ fs, which eventually shifts towards $E_F$ with time. At $\Delta t=+300$ fs, the UHB lies across $E_F$ and cannot be spectrally resolved as shown in Fig.~\ref{Figure3_PP_3}(e) (right panel). All the observed characteristics of the dispersive band have a close resemblance to the band structure of the unreconstructed metallic phase about $\Gamma$\cite{ch6_sohrt2014}. Therefore, the above results demonstrate two features near $E_F$: (i) depletion of the LHB intensity and emergence of a dispersive band above $E_F$ and (ii) shift of the UHB towards $E_F$ indicating a reduction of the Coulomb repulsion strength\cite{ch6_sohrt2014,ch6_kaiser2014}. The former corresponds to the relaxation of the PLD towards the undistorted high-temperature (metallic) phase, whereas the latter indicates photoinduced modification of the Mott-Hubbard gap. These provide evidence for phase coexistence in 1$T$-TaS$_2$ under non-equilibrium conditions, which might arise due to a particular lattice structure comprising hexagonal or striped SD domains separated by metallic islands. The manifestation of such a lattice configuration in the electronic band structure can be addressed through ARPES studies on the nearly commensurate and triclinic CDW phases of 1$T$-TaS$_2$. Altogether, our trARPES results at early time delays show that under the destruction of electronic and lattice order in 1$T$-TaS$_2$, it enters a transient phase that has remarkable similarities with the unreconstructed metallic phase, along with coexistence of the metallic (high-temperature) and insulating (Mott) phases.
 
\begin{figure}[t]
\centering
%\vspace{-4ex}
\includegraphics[width = 1\linewidth]{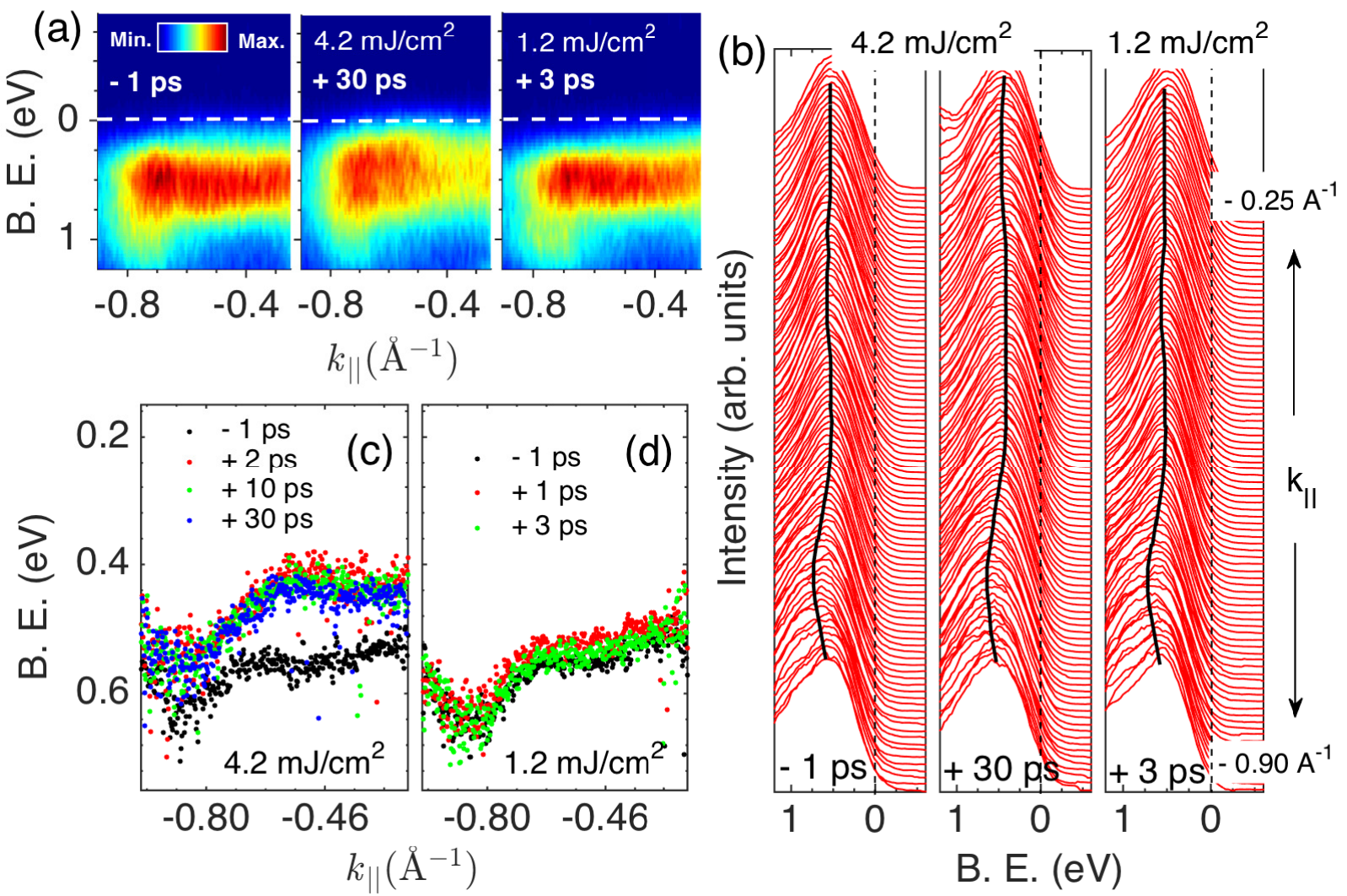}

%\vspace{-2ex}
%\vspace{-2ex}
\caption{(a) ARPES snapshots of the Ta $5d$ subband at 0.5 eV below $E_F$ along $\Gamma M$ direction taken before and after photoexcitation: delay $\Delta t$ = - 1 ps (left), $\Delta t$ = + 30 ps at pump fluence 4.2 mJ/cm$^2$ (middle), $\Delta t$ = + 3 ps at pump fluence 1.2 mJ/cm$^2$ (right). (b) Corresponding stacked EDCs representing the band dispersion. Smooth black curves are guide to the eye for the dispersion and dashed lines denote $E_F$. (c) Peak positions of the EDCs plotted as a function of $k_{||}$ at various time delays for high fluence, 4.2 mJ/cm$^2$. (d) The same for low fluence, 1.2 mJ/cm$^2$. The data at high fluence shows the presence of a long-lived state. Binding energy is abbreviated to B. E.}
\label{Figure4_PP}
%\vspace{-2ex}
\end{figure}

Now, we will move on to the recovery dynamics and identify the nature of the phase where it settles at longer time delays. Figure~\ref{Figure2_PP} captures such dynamics under strong photoexcitation (3.6 mJ/cm$^2$) for the probed Ta $5d$ subbands ($B_2$ and LHB). We observe that, as the relaxed lattice structure of the transient phase starts to recover after 300 fs, there is only a partial recovery of the lattice order till $\Delta t=+600$ fs shown in Fig.~\ref{Figure2_PP}(a). We call it partial since $B_2$ band does not exhibit the dispersion corresponding to that of before photoexcitation ($\Delta t=-1.2$ ps). Any further recovery occurs on extremely long time scales which can be clearly identified from the negligible changes in the band dispersion from 600 fs to $\Delta t=+3.5$ ps. This signifies the emergence of a long-lived metastable state in photoexcited 1$T$-TaS$_2$. The ARPES snapshots taken before and after (3.5 ps) photoexcitation, and their EDCs are shown in Fig.~\ref{Figure2_PP}(b) and Fig.~\ref{Figure2_PP}(c), respectively. In the long-lived hidden phase, $B_2$ exhibits a weaker band dispersion in comparison to the transient phase [compare red and yellow curves in Fig.~\ref{Figure2_PP}(a)]. However, the band minima is still shifted by $\approx 0.08$ eV towards $E_F$ and $B_2$ has a larger bandwidth with respect to the ground state dispersion [see Fig.~\ref{Figure2_PP}(d)]. On the other hand, the dynamics of the LHB display a complete recovery of the Mott phase. This can be claimed from the following features of LHB at $\Delta t=+3.5$ ps: (i) the spectral weight recovery in LHB and no additional weight at the tail of the EDC in Fig.~\ref{Figure2_PP}(f), and (ii) the peak of the EDC lying at a similar binding energy as that of before photoexcitation [see Figs.~\ref{Figure2_PP}(e) and \ref{Figure2_PP}(f)]. However, the recovery of LHB intensity slows down after 600 fs, with no pronounced changes at longer time delays. It is not known whether such slow dynamics of LHB can be linked to the destruction of CDW order and will require fluence-dependent studies in the future to make any further comments. 

Finally, we will look at the features of the metastable state under strong (4.2 mJ/cm$^2$) and weak (1.2 mJ/cm$^2$) photoexcitation by tracking the dynamics of the Ta $5d$ subband lying at 0.5 eV below $E_F$ (we call it $B_1$)\cite{ch6_sohrt2014} in Fig.~\ref{Figure4_PP}. For a high photoexcitation strength, the band dispersion at long time delays is stronger and shifted towards $E_F$ while this is not the case at a low photoexcitation strength [see Figs.~\ref{Figure4_PP}(c), (d)]. We show the data at $\Delta t=+30$ ps for pump fluence 4.2 mJ/cm$^2$ in Figs.~\ref{Figure4_PP}(a), (b) to emphasize that the dispersion (CDW lattice order) has not recovered even at longer times. The quantitative changes in the band structure at $\Delta t=+2$ ps are persistent till $\Delta t=+30$ ps and longer under strong photoexcitation in Fig.~\ref{Figure4_PP}(c). This, once again, provides evidence that the system is driven to a long-lived metastable state prior to the complete recovery of the CDW lattice order. On the contrary, we do not find any signatures of the metastable state under weak photoexcitation since the small bandshifts are completely recovered within $\Delta t=+3$ ps [compare black and green curves in Fig.~\ref{Figure4_PP}(d)]. At low photoexcitation strengths ($\sim1.3$ mJ/cm$^2$ in this study), the LHB dynamics show complete recovery of the Mott phase. Hence, the long time dynamics of the Ta $5d$ subbands (LHB, $B_1$, $B_2$) provide insights into the metastable phase in 1$T$-TaS$_2$, which is a hidden phase having no counterparts in equilibrium.

\section{DISCUSSION}

The correspondence between the (photoinduced) transient and (equilibrium) structurally undistorted phases imply that the ordering in the CCDW-Mott phase is destroyed as the lattice order relaxes to the undistorted metallic phase. Although the recovery of both the CDW and Mott phases begin at the same time, the CDW phase undergoes only a partial recovery while the Mott phase fully recovers within one ps. The metastable state attained by the system after its partial recovery does not correspond to any of the thermally accessible equilibrium phases. The signatures of the metastable phase are exhibited only by  $B_1$ and $B_2$ bands, while LHB shows no evidence of such a long-lived state. Since the LHB is derived from electron-electron interactions and $B_1$ and $B_2$ have dominant contribution from electron-lattice interactions\cite{ch6_smith2006}, it can be inferred that it is primarily the interaction of the electrons with the lattice that pushes the material towards a long-lived state. Such a state could be mediated by mode-selective electron-phonon coupling due to the destruction of CDW order, as has been shown for a similar compound 1$T$-TaSe$_2$\cite{ch6_shi2019}. %This follows from the fact that, once the CDW order is completely melted above a critical pump fluence, the coupling of the electrons with the lattice switches from being nearly homogenous to mode-selective

It is the electronic and lattice configuration in the low-temperature CCDW phase, which makes it susceptible to a Mott-Hubbard transition. Even though the CDW phase is not observed to reform completely, the ordering of the electronic and lattice degrees of freedom is such that the intracluster Coulomb repulsion ($U$) is larger than the electronic hopping strength ($W$), i.e., $U/W\gtrsim1.2$\cite{ch6_perfetti2008}. This tends to localize the electrons at the atomic sites, leading to the recovery of the Mott phase. Therefore, it can be deduced that the metastable state is indeed a Mott insulating phase but with a reduced CDW amplitude as compared to the CCDW phase in equilibrium. A clear and direct investigation of the structural configuration in the metastable non-equilibrium phase can be obtained from time-resolved electron diffraction, which will be used in future studies to probe the long-lived hidden phases in this compound. It is also important to identify the critical fluence above which such a long-lived hidden phase emerges. Further time-resolved studies in this direction would involve a deeper investigation of how the microscopic interactions evolve as the material changes its state under non-equilibrium conditions.

\section{CONCLUSION}
In summary, we demonstrated the characteristics of the non-equilibrium phases in photoexcited 1$T$-TaS$_2$ using time-resolved ARPES. In the transient phase, the Mott-CDW order is suppressed and the band structure has excellent resemblance with that of the unreconstructed metallic phase. Together with the complete relaxation of the PLD driven by charge redistribution, the dynamics at early time delays also exhibit signatures of phase coexistence in photoexcited 1$T$-TaS$_2$. The Mott and CDW orders begin recovering around the same time, but only to settle in a long-lived metastable phase. In this ``hidden" phase, 1$T$-TaS$_2$ is a CCDW-Mott insulator but with a reduced CDW amplitude and the emergence of this phase is driven by the lattice order. In addition, the metastable state emerges only under strong photoexcitation of the system. A distinct characterization of these phases provides deeper insights into the state of charge and lattice order under non-equilibrium conditions and the prominent role played by the different degrees of freedom in governing these phases in a complex system.

% Thus, trARPES and other pump-probe techniques provide the possibility to uncover previously unknown phases in several complex systems by controlling the external parameters.

%We also provided further perspectives for a better characterization of these phases, which might make this material more suitable for the fabrication of new devices.

\section*{ACKNOWLEDGEMENTS}
We are thankful to E. Nicolini and G. Bortoletto for characterization of the samples using Laue diffraction. We acknowledge fruitful discussions with M. Capone, Z. Bacciconi and A. Amaricci. This work was supported by the FLAG-ERA grant DIMAG, by the Research Foundation – Flanders (FWO), the Agence Nationale pour la Recherche (ANR), the Deutsche Forschungsgemeinschaft (DFG), the Slovenian Research Agency (ARRS).

%We are thankful to ELETTRA synchrotron facility for proving us an access to the BaDElPh beamline that contributed to the results presented in this work. We thank L. Sancin for technical assistance during experiments and J. Mravlje and F. Galdenzi for fruitful discussions during the beamtime.

%\section*{Supplementary Material}

%\beginsupplement

%\subsection{Experimental geometry of the time-resolved photoemission setup}
%\subsection{Photoinduced dynamics of the Ta $5d$ subband at 0.5 eV below E$_F$}

%\subsection{}

%\subsection{Comparison of photon energy dependence of valence band peak position among Ta$_2$Ni(Se$_{1-x}$S$_x$)$_5$ compounds}

\end{document}